\documentclass[traditabstract]{aa} % for the abstract without structuration 
\usepackage[dvips]{graphicx}          % this allows the inclusion of figures
\usepackage[comma,authoryear]{natbib} % this allows the use of natbib
\bibpunct{(}{)}{,}{a}{}{,}            % this alters the punctuation used by natbib

\def\note #1]{{\rm #1]}}
\def\Msun{\,{\rm M}_\odot}
\def\dd{{\rm d}}

\begin{document}

   % place your title here:
   \title{Open issues in stellar modelling}

   % place an option sub-title here:
   %\subtitle{I. Apparition of new sunspots as observed by SOHO}

   % put the names of the authors:
   \author{J{\o}rgen Christensen-Dalsgaard
          }

   % put the name and addresses of the institutes
   \institute{Department of Physics and Astronomy, Aarhus University,
8000 Aarhus C, Denmark;
              \email{jcd@phys.au.dk}
             }

  % place your abstract here:
  \abstract
{An important goal of helio- and asteroseismology is to improve the modelling
of stellar evolution.
Here I provide a brief discussion of some of the uncertain issues in
stellar modelling, of possible relevance to asteroseismic inferences.}

   % keywords go here:
   \keywords{ solar interior -- helioseismology -- stellar evolution --
physics of stellar interiors -- asteroseismology}

   \maketitle
   
%-----------------------------------------------------------------------
% the main body of the article:

\section{Introduction}

The goal of the workshop was to investigate ways to improve our understanding
of the Sun;
this is obviously intimately linked to the general understanding of stellar
structure and evolution, and indeed there are considerable prospects that
our growing possibilities of asteroseismic sounding of other stars will 
inform our studies of the solar interior.

From the point of view of asteroseismology, the relevant aspects of stellar
modelling include both the study of stellar structure and evolution and
the modelling of stellar oscillations, in particular their frequencies, 
for a given model. 
The latter aspect provides the diagnostic link between the observations and
the stellar models;
although the adiabatic approximation is valid for the oscillations
in most of the star,
departures from adiabaticity, and other uncertainties in the modelling of 
the near-surface layers, give rise to substantial systematic errors that 
must be taken into account in the analyses.

Unlike what is perhaps a common perception, we are still far from adequate
modelling of stellar interiors. 
Here I can only touch on a few issues, mainly in connection with the modelling
of main-sequence stars showing solar-like oscillations.
A more detailed discussion of these issues, and further references,
was provided by \citet{Christ2010}.
For an extensive presentation of stellar oscillations and helio- and 
asteroseismic techniques and results, see \citet{Aerts2009}.

\section{Numerical issues}

A prerequisite for meaningful asteroseismic diagnostics of the physics
of stellar interiors is that stellar modelling presents a faithful
representation of the physical assumption.
Thus the models must be numerically sufficiently accurate.
Perhaps the best test of this is to compare independently computed models
with the same physical assumptions.
A major effort towards such comparisons was carried out in the ESTA 
project initiated as part of the preparations for analysis of CoRoT data
\citep{Montei2008}.
In general, the agreement was reasonable between the results of
the evolution codes included in the comparison, although not obviously
adequate for the asteroseismic analysis;
also, it would be very valuable to extend the comparison to other codes
commonly used for general stellar modelling.
The results of adiabatic oscillation calculations for a given model 
agreed quite well, but the analysis highlighted the importance of adequate
numerical resolution in the evolution and oscillation calculations, and
of consistency in the description of the stellar physics.

An important, and probably often inadequately treated, 
aspect of the computations
is the accurate specification, implementation and documentation
of the unavoidable approximations in stellar modelling.
Without appropriate care in this area it is difficult or impossible to 
use asteroseismic inferences to test the validity of these approximations.

\section{Stellar parameters}

%\note [Stress importance of knowing as much as possible about the stars
%being modelled. Teff, L, log g, for example.]
%
Efficient utilization of the asteroseismic data requires the best possible
information about other properties of the star.
In the solar case, the mass, radius, luminosity and age are determined
quite accurately (or at least precisely) from independent observations.
Solar composition, characterized by the ratios of abundances of elements
heavier than helium to the abundance of hydrogen, can be determined from 
spectroscopic observations.
Recently, however, there has been a substantial revision in some of these
abundances, leading to conflicting comparisons between the resulting
solar models and helioseismic inferences
\citep[e.g.,][and references therein]{Asplun2009}.

Parameters of other stars are in general known far less well.
Masses can be obtained in the rare cases where the star is a member of a
well-observed binary system. 
The effective temperature and surface gravity can be obtained 
from spectroscopic observations, but subject to the possible limitations
of modelling of stellar atmospheres and hence with substantial 
(and probably often underestimated) uncertainties.
The stellar luminosity requires knowledge, from parallax observations, of
the distance and hence is currently restricted to relatively nearby stars;
also, to the observed stellar magnitude must be applied a bolometric 
correction which again depends on atmosphere models.
Also, in a few cases the stellar radius can be determined from interferometry,
again assuming that the distance is known and with some sensitivity to
atmospheric structure through limb darkening.
Stellar composition is obtained from spectroscopy;
in the case of stars similar to the Sun this is most often done differentially,
relative to the solar spectrum, and hence the abundances are directly 
affected by the uncertainty in the solar composition.

\section{Microphysics}

%\subsection{Equation of state}

For the purpose of stellar modelling the equation of state is probably
in general sufficiently well known from recent tabulations of sophisticated
equations of state.
On the other hand, helioseismology clearly demonstrates that these are not
yet correct, at the level of the observational precision
\citep[e.g.,][]{Basu1999}.
This would also affect the asteroseismic determination of the helium abundance
from the signatures in the frequencies of helium ionization \citep{Houdek2007}.

%\subsection{Opacity}

The computation of stellar opacities is considerably more uncertain
than the equation of state, with direct effect on the structure of the
radiative part of stellar models.
Since the heavy-element abundance affects stellar structure predominantly
through the opacity, uncertainties in the heavy-element abundances and the
opacity are closely linked. 
Thus an obvious way to correct solar models, given the revised abundances,
is to claim substantial opacity increases 
\citep[e.g.,][]{Bahcal2005, Christ2009}, although possibly beyond what is
physically realistic.
An independent indication of a need for opacity increases, although at somewhat
lower temperatures than relevant in the Sun,
comes from the lack of predicted instability of some observed modes in
$\beta$ Cephei stars \citep{Dziemb2008}.

%\subsection{Nuclear reactions}

Although there remain substantial uncertainties in important nuclear parameters
the effect on stellar modelling is in general relatively modest, since
even large changes in the parameters can be compensated by modest changes
in the temperature, owing to the high temperature sensitivity of the reactions.
An important exception concerns the balance between contributions to the
PP chains and the CNO cycle in hydrogen burning, which has a substantial
effect on the presence and extent of convective cores.
This includes the relatively recent large reduction in the rate of
proton capture by ${}^{14}{\rm N}$ \citep{Angulo2005}.
An interesting, and so far not resolved, issue concerns electron
screening of nuclear reactions \citep{Shaviv2004, Mao2009}.

%\subsection{Diffusion and settling}

%\note [Element-dependent effects of radiative levitation. Greatly complicates
%modelling (and is most often ignored).]

There is no doubt that diffusion and settling
take place in those parts of a star where there is no macroscopic
motion. 
These processes, in various approximations, are now universally included in
`standard' solar modelling, leading to a increase of a few per cent in the
surface hydrogen abundance during evolution to the present solar age, and
a decrease of around 10 per cent in the heavy-element abundances.
In somewhat more massive stars with thinner outer convection zones the
settling rate at the base of the convection zone is much higher, leading
to an almost complete elimination at the stellar surface
of helium and heavier elements, on a timescale
short compared with the evolution timescale \citep{Vaucla1974}.
To account for the `normal' abundances observed in most such stars one must
therefore invoke mixing processes or possibly mass loss to compensate for
the settling.
A possible explanation is mixing caused by rotationally induced meridional
circulation (see below).

An additional complication, particularly in stars a little more massive
than the Sun, is the selective effects of radiation pressure on different
elements, leading to gravitational levitation counteracting settling 
and strong local variations in the heavy-element composition.
To be taken properly into account, this requires opacity calculations
for the local composition, depending on location and time, as the star
evolves \citep[e.g.,][]{Richer2000}.
This has so far only been consistently implemented in very few evolution
calculations.

%\note [Too rapid depletion in stars with thin
%outer convection zones.
%Must return to that in connection with rotation.]

\section{Properties of convective cores}

%\note [Effects of microphysics on convective cores.]
%
Convective cores play an important role in the main-sequence evolution
of stars of masses just slightly higher than the Sun and above.
This is caused by the increasing dominance in hydrogen burning
of the much more temperature sensitive CNO cycle over the PP chains.
For stars of masses less than around 2 solar masses this involves a phase
where the mass of the convective core increases, owing to the gradual
conversion, on a timescale comparable to the evolution timescale, of
${}^{16}{\rm O}$ to ${}^{14}{\rm N}$;
if diffusion is neglected
the growth of the core leads to a discontinuity in the hydrogen abundance
at the edge of the core, and hence in the density and sound speed.
Asteroseismic diagnostics of this discontinuity, and other aspects
of convective cores, may be possible with sufficiently accurate data
\citep{Popiel2005, Mazumd2006, Cunha2007}.

The uncertainties in the microphysics, particularly as it affects the 
importance of the CNO cycle, influence the size of the convective core
\citep[see][for details]{Christ2010}.
Thus the reduction in the ${}^{14}{\rm N}$ reaction rate shifts the onset
of convective cores higher in stellar mass by about $0.06 \Msun$.
An interesting case is the effect of the revision of solar abundances
which, as discussed above, is reflected in the assumed stellar abundances.
\citet{Vanden2007} showed that this led to a significant change in the
isochrones computed for the open cluster M67;
with the old composition models near the end of the central hydrogen burning
had a convective core, as also suggested by the observed colour-magnitude
diagram, while models with the revised composition lacked the convective core.
%\note [Could illustrate, if there is room.]

%\note [Convective core overshoot.]

A probably more important uncertainty concerns the extent of convective
overshoot.
There is little doubt that motion continues beyond the convectively 
unstable region, but the extent of that motion, and its effects on
stellar structure, are highly uncertain.
Presumably the motion is sufficiently vigorous to cause homogenization
of the composition, but it is less clear whether it leads to full mixing
of entropy and hence an adiabatic stratification.
The extent is typically parameterized as a fraction $\alpha_{\rm ov}$
of the pressure scale height at the edge of the core, 
with a correction for very small cores, but no {\it a priori} estimate
of $\alpha_{\rm ov}$ is available.
Analyses of open clusters and binary stars lead to values of $\alpha_{\rm ov}$
of typically around 0.1 -- 0.2, 
a value confirmed by asteroseismic analyses of $\beta$ Cephei stars
\citep[e.g.,][]{Aerts2003}.
   
   \begin{figure}[htbp]
     % This figure is 8 cm wide:
     \begin{center}
       \includegraphics[width=8cm]{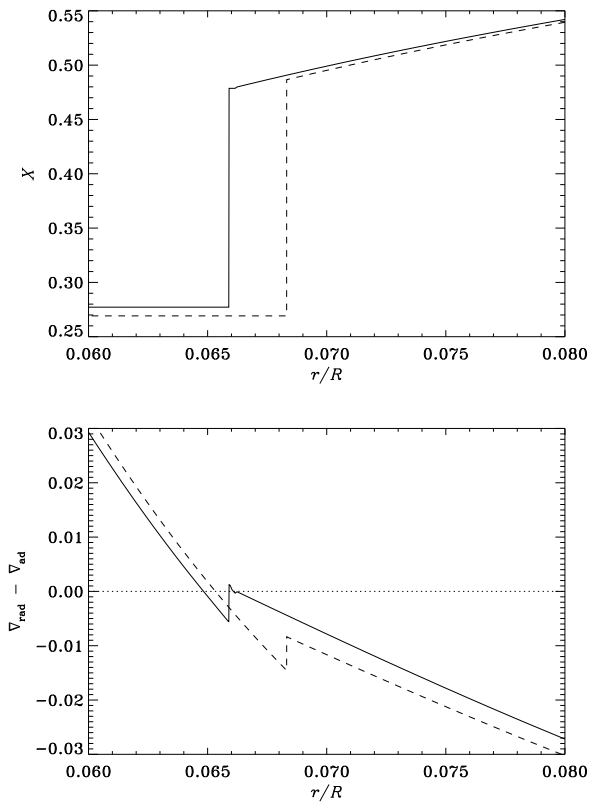}
     \end{center}
     \caption{Properties near the edge of the convective core
in models of a $1.3 \, \Msun$ star,
as a function of distance to the centre in units of the stellar radius;
solid lines show a model without overshoot, dashed lines a model with modest
overshoot, $\alpha_{\rm ov} = 0.05$. 
The upper panel shows the hydrogen abundance by mass, illustrating the
discontinuity at the edge of the mixed core.
The lower panel shows the departure of the radiative temperature gradient 
$\nabla_{\rm rad}$ from the corresponding adiabatic gradient.}
     \label{fig:grad}
   \end{figure}

%\note [Briefly on semiconvection; here probably include a figure, ideally 
%with overshoot also.]

An additional complication in models with growing convective cores is
the presence of what has been called semi-convection.
Convective instability is typically defined in terms of
the temperature gradient $\nabla = \dd \ln T / \dd \ln p$, where
$T$ is temperature and $p$ is pressure;
convective instability sets in where the value $\nabla_{\rm rad}$ of $\nabla$
required to transport energy by radiation exceeds the adiabatic value
$\nabla_{\rm ad}$.
Since $\nabla_{\rm rad}$ is proportional to the opacity which increases with
increasing hydrogen abundance $X$, 
$\nabla_{\rm rad}$ jumps to a larger value at the
edge of a growing convective core where $X$ is discontinuous.
This is illustrated in Fig. 1, in a model calculation where the 
extent of the convective core is defined with marginal instability
evaluated for the composition outside the core, 
leading to a small convectively stable region in the outer parts of the core.
Treating the mixed region in this manner is obviously a computational artifice,
rather than a physically justified model.
Modelling of this region, with added complications
when the settling of heavy elements is taken into account,
tends to lead to irregular fluctuations in the
mass contained in the convective core \citep{Lebret2008}.
As illustrated, even a modest amount of overshoot shifts the composition
discontinuity sufficiently far into the stable region that the problem is
eliminated.

%\note [Could add a little on effects of overshoot on this, if there is time!]

%\note [Very briefly on saltfinger instability, perhaps.
%Probably not!]

\section{Near-surface problems}

%\note [Convection, obviously, and possibility of calibrating with 
%hydrodynamical simulations.]
%
The treatment of convective envelopes also involves substantial uncertainties.
Overshoot below the convective envelope has a relatively modest effect on
stellar evolution although it can affect the properties of the red bump
on the red-giant branch. % \note [A reference would not hurt.]
A more serious concern are the properties of the near-surface layer where
the density is low and consequently a substantial superadiabatic gradient
is required to transport the energy.
Together with the structure of the stellar atmosphere this determines the
specific entropy in the predominantly adiabatic bulk of the convective
envelope and hence its structure, including its depth.
In the solar case the treatment of this layer, e.g., using the
\citet{Bohm1958} mixing-length formulation, is calibrated to obtain
the correct radius;
this calibration is typically, with little justification, used in modelling
other stars.
Hydrodynamical simulations of near-surface convection \citep{Nordlu2009}
provide a reasonably realistic modelling of these layers;
unlike other parts of the star the relevant dynamical and thermal timescales
are sufficiently similar that the relevant effects can be taken into
account, 
although obviously still with an approximate treatment of scales smaller
than the numerical resolution.
The results of the simulations can then be used to calibrate the simpler
formulations \citep[e.g.,][]{Trampe2007};
this offers a promising procedure for more realistic modelling of this
part of the star, although it has so far not seen much use.

Uncertainties in the modelling of the near-surface layers have a substantial
effect on the oscillation frequencies and hence on their use as asteroseismic
diagnostics.
In addition to the structure of the superadiabatic layer, these uncertainties
include the dynamical effects, usually ignored,
of convection on stellar structure in the form of `turbulent pressure',
nonadiabatic effects on the oscillations,
and the coupling between convection and pulsations, in terms of 
the perturbation to the convective flux and the turbulent pressure,
as well as the stochastic excitation of the modes, for solar-like oscillations.
These effects dominate the difference between the observed and modelled 
frequencies of solar oscillations;
they can be suppressed, however, in helioseismic analyses because of the
broad range of degrees of the observed modes.
In the stellar case this is not possible, in general.
It was pointed out by \citet{Roxbur2003} that combinations of 
frequency separations can be constructed which are insensitive to the
superficial layers and retain their sensitivity to the properties of the core
\citep[see also][]{Oti2005}.
For more general use of the frequencies, including calibrations of the
overall properties of the star, one can attempt to estimate the 
near-surface effects on the frequencies, by assuming a functional form
similar to the known effect in the solar case \citep{Kjelds2008}.
This, however, remains a serious issue in asteroseismic analyses.

%\note [Other uncertainties in modelling: nonadiabatic effects, 
%pulsation/convection interaction.]

%\note [Very briefly on taking it out.]

\section{Rotation}

%\note [Dynamical effects; relatively straightforward, at least when rotation
%is not too fast. 2D modelling.]
%
There is no doubt that most, or indeed all, stars rotate, yet rotation
is usually ignored in modelling of stellar evolution.
A detailed discussion of the effects of rotation on stars was recently
provided by \citet{Maeder2009}.

The dynamical effects of rotation on stellar structure
are relatively straightforward to
incorporate, at least as long as they can be treated as perturbations
around a non-rotating, spherically symmetric structure.
In a slowly rotating star such as the Sun these effects are very small.
However, many stars rotate so rapidly that the perturbative approach
is inadequate;
here two-dimensional modelling of stellar structure is required
\citep[e.g.,][]{Roxbur2004, MacGre2007}.
Far greater complications are associated with the effects on stellar 
evolution, including the evolution of the internal rotation rate.
A naive local application of the conservation of angular momentum
would predict that the angular velocity of the central parts of stars 
will increase with age as these regions contract, while rotation in the outer
parts would be expected to slow down as they expand.
This is certainly too simple.
As already noted by von Zeibel and Eddington, rotation causes a thermal
imbalance which leads to circulation and hence redistribution of angular
momentum and mixing of the stellar composition.
Indeed, it is likely that in many stars this mixing counteracts the 
rapid settling discussed above.
To these processes must also be added mass loss, possibly magnetically
linked to the stellar convective envelope, which removes angular momentum
from the star.
It seems likely that most stars start their life with rapid rotation;
stars with masses up to somewhat higher than the Sun apparently
lose angular momentum to a magnetized stellar wind, leading to a strong
decrease in rotation with age \citep[e.g.,][]{Barnes2003}.

A treatment of these processes was proposed by \citet{Zahn1992}
and further developed by \citet{Maeder1998}.
This assumes an angular velocity that depends only on the distance to the
centre of the star, as a result of strong horizontal turbulence.
Mixing of composition is a diffusive process while transport
of angular momentum in addition includes advective terms.
This formulation has seen fairly extensive use and has had some
success in accounting for the observed composition of massive stars.

A serious problem is to account for the helioseismically inferred
solar internal rotation rate \citep[e.g.,][]{Howe2009};
in particular, the Zahn model is unable to explain the present slow
rotation of the radiative interior.
This requires additional mechanisms transporting angular momentum
from the interior to the convection zone.
It has been proposed that this coupling could be mediated by gravity
waves excited at the base of the convection zone \citep{Mathis2008};
alternatively, it may be of magnetic nature \citep{Garaud2009}.
It is obvious that asteroseismic information about the internal rotation 
of other stars,
although unavoidably quite limited in the foreseeable future, 
can be extremely valuable in distinguishing between these mechanisms.

%\note [Effects of transport and mixing. Angular-momentum evolution.]

%\note [Effects of magnetic fields. What controls the solar internal rotation.]

\section{Concluding remarks}

%\note [Need to investigate the seismic signatures of some of the physical
%effects discussed here; and how to identify and probe them.]
%
It is evident that there are many serious open issues in stellar modelling.
An important task is the evaluation of the asteroseismic
signatures of these effects, including the design of diagnostics that
may best investigate them and a determination of the resulting
requirements on the observations.
The asteroseismic observations that are currently been obtained by the
CoRoT and Kepler space missions certainly provide excellent prospects for
addressing these issues, although the experience from CoRoT has shown
that the analysis of the data also involves serious challenges.

There is clearly a need for very substantial development of the techniques
of stellar modelling. 
This can be inspired, but certainly not replaced, by further detailed 
hydrodynamical simulations of specific aspects of stellar interior dynamics.
Except for the near-surface layers a serious constraint is the huge mismatch
between the relevant dynamical and thermal timescales, implying that the
simulations cannot be run under realistic stellar conditions.
A great deal of physical insight will therefore be required to extrapolate
the results of the simulations to those conditions.

These efforts will require intensive collaborations between data analysis,
data interpretation, theory and modelling.
A workshop such as the present is an ideal venue for furthering such
collaboration and create new ideas.

\begin{acknowledgements}
I am very grateful to the organizers of the Ponte de Lima workshop,
in particular Margarida Cunha, for an extremely productive and enjoyable
meeting.
Participation in the workshop was supported by the European
Helio- and Asteroseismology Network (HELAS), a major international collaboration
funded by the European Commission's Sixth Framework Programme.
\end{acknowledgements}

%-----------------------------------------------------------------------
% the bibliography (using natbib):

%-----------------------------------------------------------------------
\end{document}